\documentclass[12pt]{iopart}
\usepackage{iopams}  
\input epsf
\expandafter\let\csname equation*\endcsname\relax
  \expandafter\let\csname endequation*\endcsname\relax
  \usepackage[fleqn]{amsmath}
\usepackage{amsfonts}
\usepackage{amssymb}
\usepackage{amsopn}
\usepackage{epsfig}
\usepackage{graphics,psfrag,rotating}
\usepackage{graphicx}
\usepackage{dcolumn}
\usepackage{bm}
\usepackage{epstopdf}
\usepackage{color}
\usepackage[usenames,dvipsnames,svgnames]{xcolor}
\usepackage[colorlinks=true,
            linkcolor=red,
            urlcolor=gray,
            citecolor=blue]{hyperref}
  \usepackage{hyperref}
\def \D {\tilde{\nabla}}

\def \ep {\varepsilon}

\def\md{\mathcal{D}}
\def\mz{\mathcal{Z}}

\def\k{\kappa}

\def\3nab{\tilde{\nabla}}

\def\tl{\tilde}
\def\hsp5{\hspace{5mm}}
\newcommand{\sfrac}[2]{{\textstyle{#1\over#2}}}
\def\case#1/#2{\textstyle\frac{#1}{#2}}

\def\ber {\begin{eqnarray}}
\def\eer {\end{eqnarray}}
\def\bea {\begin{eqnarray}}
\def\eea {\end{eqnarray}}

\def\bc {\begin{center}}
\def\ec {\end{center}}
\def\case#1/#2{\frac{#1}{#2}}

\newcommand{\nn}{\nonumber\\}

\newcommand{\del}{\Delta}
\newcommand{\be}{\begin{equation}}
\newcommand{\bse}{\begin{subequation}}
\newcommand{\ese}{\end{subequation}}
\newcommand{\ee}{\end{equation}}
\newcommand{\eei}{\end{eqnarray}\indent\indent}
\newcommand{\ba}{\begin{array}}
\newcommand{\ea}{\end{array}}
\newcommand{\bal}{\begin{eqnarray}}
\newcommand{\eal}{\end{eqnarray}}

\newcommand{\hs}{\,-\,}

\newcommand{\Lda}{\Lambda}
\newcommand{\Om}{\Omega}

\newcommand{\car}{{\cal R}}

\def\case#1/#2{\textstyle\frac{#1}{#2} }
\newcommand{\nb}{\nabla}

\newcommand{\gd}{g_{ab}}
\newcommand{\bver}{\begin{verbatim}}
\newcommand{\ever}{\end{verbatim}}

\newcommand{\tlnb}{\tilde{\nabla}^{2}}
\newcommand{\np}{\newpage}

\begin{document}
\title{Breaking the cosmological background degeneracy by two\hs fluid perturbations in $f(R)$ gravity}
%%%%%%%%%%%%%%%%%%%%%%%%%%%%%%%%%%%%%%%%%%%%%%%%%%%%
\author{Amare Abebe \footnote{
amare.abbebe@gmail.com}}
\address{Department of Physics, North-West University,  Mahikeng 2735, South Africa.}
\address{Entoto Observatory and Research Center, P.O.Box 33679, Addis Ababa, Ethiopia.}

%%%%%%%%%%%%%%%%%%%%%%%%%%%%%%%%%%%%%%%%%%%%%%%%%%%%%
\begin{abstract}
One of the exact solutions of  $f(R)$ theories of gravity 
in  the presence of different forms of matter exactly mimics the $\Lambda$CDM solution 
of general relativity at the background level. In this work we study the evolution of scalar 
cosmological perturbations in the covariant and gauge\hs invariant formalism and show that although the background in such a model is indistinguishable from the 
standard $\Lambda$CDM cosmology, this degeneracy is broken at the level of  first\hs order perturbations.  This is done by predicting 
different rates of structure formation in $\Lda$CDM and the $f(R)$ model both in the complete and quasi\hs static regimes. \end{abstract}
%%%%%%%%%%%%%%%%%%%%%%%%%%%%%%%%%%%%%%%%%%%%%%%%%%%%%
  
 \maketitle

%%%%%%%%%%%%%%%%%%%%%%%%%%%%%%%%%%%%%%%%%%%%%%%%%%%%%

\section{Introduction}\label{intro}
The $\Lambda$CDM (or {\em Concordance}) Model  of cosmology \cite{ostriker95} is one of the 
greatest successes  of General Relativity (GR). It reproduces beautifully all the 
main observational results \cite{dunsby10} such as the dimming of type Ia Supernovae \cite{riess98, perlmutter99, tonry03, knop03}, 
Cosmic Microwave Background  (CMB) radiation anisotropies \cite{spergel03, spergel07}, Large 
Scale Structure formation \cite{tegmark04, seljak05, cole05}, baryon acoustic oscillations \cite{sdss05} and weak 
lensing \cite{jain03}. But the $\Lda$CDM model has many serious shortcomings as well, most notably the fine-tuning problem associated with the so-called  {\it cosmological constant} and {\it coincidence} problems \cite{weinberg89}.

Among the leading  alternatives to the  $\Lambda$CDM paradigm are $f(R)$ models of gravity. These models  are based on modifications of the standard Einstein-Hilbert action and naturally
admit the currently assumed expansion history of the  Universe such as the early inflationary epoch \cite{staro80} and  the late\hs time accelerated expansion. Several recent lines of research  have therefore focused  on the viability of these theories as alternatives to dark energy and their cosmological and astrophysical applications \cite{carroll04, carroll05, hu07, bean07, bert06, Abebe2011, abebe14, nzioki14} (see the recent reviews \cite{shin07, capozziello2008extended, de2010f, sotiriou10} and
references therein for more examples) but such studies come at the cost of high complexity of the physics involved.

One approach in this direction is 
the technique of cosmological  {\it reconstruction} where it is assumed that the expansion history of
the Universe is known exactly, and the field equations are inverted to
deduce what class of theories will give rise to this particular
cosmological evolution \cite{dunsby10, goheer2009power, nojiri2010, cognola10, nojiri2007modified, nojiri06, nojiri2006dark, shin2010inverse, carloni2012}. 
In \cite{dunsby10} it was shown that the  $\Lambda$CDM expansion history can be mimicked exactly by an $f(R)$ model that describes  
 a universe filled with a minimally-coupled  and non-interacting,
massless scalar field and dust-like matter.  This means that if one has to  discriminate between  these two models, one has to go beyond  the level of the Friedmann-Lema{\^i}tre-Robertson-Walker (FLRW) background and study how the perturbations of matter in these models grow.
If the predicted rates of  structure formation appear to be different, then that is one way of breaking the background degeneracy that exists between $\Lda$CDM and the mimicking $f(R)$ model.

%%%%%%%%%%%%%%%%%%%%%%%%%%%%%%%%%%%%%%%%%%%%%%%%%%%%%%%%%%%%%%%%%%%%%%%%%%%%%%%%%%%%%%%%%%%%%%%%%%%
The rest of the  paper is organized as follows:  in Sec.\ref{backg} we give a summary of the background cosmological evolution as determined by the Concordance Model. We then give a reconstruction of the model of $f(R)$ gravity that gives the exact background expansion history as the Concordance Model.

  Sec.\ref{dynamics} deals with the formulation of the  covariant density, expansion and curvature  perturbations and their corresponding evolution equations, and the analysis of the matter power spectrum produced, followed by a brief discussion of the quasi\hs static approximation in Sec. \ref{quasi}.
  
Finally in Sec.\ref{disc} we discuss the results and give conclusions of the work.

Natural units ($\hbar=c=k_{B}=8\pi G=1$)
will be used throughout this paper, and Latin indices run from 0 to 3.
The mathematical operators $\nabla$ (or $;$),  $\D$ and the comma derivative $,$ represent the usual covariant derivative, the spatial covariant derivative, and partial differentation, respectively.  We use the
$(-+++)$ signature and the Riemann tensor is defined by
\begin{eqnarray}
R^{a}_{bcd}=\Gamma^a_{bd,c}-\Gamma^a_{bc,d}+ \Gamma^e_{bd}\Gamma^a_{ce}-
\Gamma^f_{bc}\Gamma^a_{df}\;,
\end{eqnarray}
where the $\Gamma^a_{bd}$ are the Christoffel symbols (i.e., symmetric in
the lower indices), defined by
\begin{equation}
\Gamma^a_{bd}=\frac{1}{2}g^{ae}
\left(g_{be,d}+g_{ed,b}-g_{bd,e}\right)\;.
\end{equation}
The Ricci tensor is obtained by contracting the {\em first} and the
{\em third} indices of the Riemann tensor:
\begin{equation}\label{Ricci}
R_{ab}=g^{cd}R_{cadb}\;.
\end{equation}
Unless otherwise stated, an overdot $^{.}$ represents differentiation with respect to cosmic time $t$ whereas  $f^{'}, f^{''}$, etc..  are shorthands for first, second, etc...derivatives  of the function $f(R) $ with respect to the Ricci scalar
\be
R=R^{a}_{a}\;,
\ee
and $f$ is used as a shorthand for $f(R)$.

\section{The background spacetime}\label{backg}
If  we write the generalized Einstein-Hilbert action  for  $f(R)$-gravitational interactions
 \be\label{fRaction}
 {\cal{A}}=\frac{1}{2}\int{d^{4}x\sqrt{-g}\left[f(R)+2{\cal{L}}_{m}  \right] },
 \ee
 and apply the variational principle of  least action with respect to the metric $g_{ab}$, we obtain a  generalization of the Einstein Field Equations (EFEs) given as
 
 \be\label{metricfes}
 G_{ab}=\frac{T^{m}{}_{ab}}{f'}+\frac{1}{f'}\left[\frac{1}{2}g_{ab}(f-Rf')+\nb_{b}\nb_{a}f'-g_{ab}\nb_{c}\nb^{c}f'\right]\equiv T_{ab} \;,
 \ee
where $T^{m}{}_{ab}$ is the usual energy\hs momentum tensor (EMT) of standard matter given by
\be T^{m}{}_{ab}\equiv\frac{2}{\sqrt{-g}}\frac{\delta(\sqrt{-g}{\cal{L}}_{m})}{\delta \gd}\;,\ee

 and the remaining terms on the right-hand  side of Eqn \eref{metricfes}
 \be
 T^{R}{}_{ab}\equiv\frac{1}{f'}\left[\frac{1}{2}g_{ab}(f-Rf')+\nb_{b}\nb_{a}f'-g_{ab}\nb_{c}\nb^{c}f' \right] \;,
 \ee  can be interpreted as the effective EMT for the \textit{curvature fluid}.
 
Because of the arbitrary function introduced in the Lagrangian of Eqn \eref{fRaction}, there is more freedom to explain cosmic acceleration (inflation and late-time acceleration)\cite{nojiri03} and large\hs scale structure formation without the inclusion of exotic matter and energy  (see \cite{staro80, carroll04, sotiriou10, clifton12, capozziello11extended, nojiri2011unified, Biswas12} and references therein).
 
 However,  not all functional forms of these models can be viable cosmological models: a wide range of them can be ruled out based on observations\hs cosmological and astrophysical\hs while others can be rejected because of theoretical pathologies.

 The EMTs of standard matter and that of the total fluid
are  conserved, i.e.,   \be T^{m;b}{}_{ab}=0,~~T^{;b}{}_{ab}=0\;,\ee
 but  the {\it effective} EMTs of matter $\tl{T}^{m}{}_{ab}\equiv\frac{T^{m}{}_{ab}}{f'}$ and  curvature $T^{R}{}_{ab}$ are not individually conserved \cite{carloni08}:
 \begin{align} 
 &\tl T^{m:b}{}_{ab}=\frac{T^{m;b}{}_{ab}}{f'}-\frac{f''}{f^{'2}}T^{m}{}_{ab}R^{;b},\\
 &T^{R;b}{}_{ab}=\frac{f''}{f^{'2}}\tl T^{m}{}_{ab}R^{;b}.
 \end{align}
For  a spatially flat, homogenous and isotropic (FLRW)  background, the matter energy\hs momentum tensor is given by 
\be T^{m}_{ab} = \mu_{m}u_{a}u_{b} + p_{m}h_{ab}+ q^{m}_{a}u_{b}+ q^{m}_{b}u_{a}+\pi^{m}_{ab}\;,\ee
where  $\mu_{m}$, $p_{m}$, $q^{m}_{a}$ and $\pi^m_{ab}$ denote  the standard matter energy density, 
pressure, heat flux and anisotropic pressure, respectively.  Here  $u^a$  is the $4$\hs velocity of fundamental observers:
\be u^{a} = \frac{dx^{a}}{dt}\;,\ee
whereas  \be h_{ab}=g_{ab}+u_au_b\ee
is the projection tensor into the tangent 3-spaces orthogonal to $u^a$.
In the standard  $1+3$-covariant approach, the 4-velocity vector $ u^{a} $ is used to define the 
\textit{covariant time derivative} for any tensor 
${S}^{a..b}_{c..d} $ along an observer's worldlines:
\be
\dot{S}^{a..b}_{c..d}{} = u^{e} \nb_{e} {S}^{a..b}_{c..d}~,
\ee
and the tensor $ h_{ab} $ is used to define the fully orthogonally 
\textit{projected covariant derivative} $\tl\nb$ for any tensor ${S}^{a..b}_{c..d} $:
\be
\tl\nb_{e}S^{a..b}_{c..d}{} = h^a_f h^p_c...h^b_g h^q_d 
h^r_e \nb_{r} {S}^{f..g}_{p..q}\;,
\ee
with total projection on all the free indices. 
The covariant derivative of the timelike vector $u^a$ is decomposed into its
irreducible parts as
\be
\nb_au_b=-A_au_b+\sfrac13h_{ab}\Theta+\sigma_{ab}+\ep_{a b c}\omega^c,
\ee
where $A_a=\dot{u}_a$ is the acceleration, $\Theta=\tl\nb_au^a$ is the expansion, 
$\sigma_{ab}=\tl\nb_{\langle a}u_{b \rangle}$ is the shear tensor and $\omega^{a}=\ep^{a b c}\tl\nb_bu_c$ 
is the vorticity vector. 

The energy conservation equation for standard matter is given by
\begin{equation}\label{cons:perfect}
\dot{\mu}_{m}=-3H\left(\mu_{m}+p_{m}\right)\;,
\end{equation}
where $H=\frac{\dot{a}}{a}=\sfrac{1}{3}\Theta$ is the Hubble expansion parameter and  $a=a(t)$ is the cosmological scale factor.
For flat FLRW spacetimes, the background cosmological expansion history is described by the Raychaudhuri equation
\bea
\label{ray}
\dot{H}+H^2&=&-\frac{1}{6f'}\left[\mu_{m}+3p_{m}+f-f'R+3H f''\dot{R}+3f'''\dot{R}^2+3f''\ddot{R}\right]\;,
\eea 
 the Friedmann equation
\begin{equation}\label{fried}
H^2= \frac{1}{6f'}\left[2\mu_{m}+Rf'-f-6H f'' \dot{R}\right]\;;
\end{equation}
%the {\em trace equation}
%\be
%\label{trace}
%3\ddot{R}f''=\mu-3p+ f'R-2f-9H f''\dot{R}-3f'''\dot{R}^2\;;
%\ee

and the equation for the Ricci scalar,

\be\label{R1}
R=6\left(\frac{\ddot{a}}{a}+\frac{\dot{a}^2}{a^2}\right)=6\dot{H}+12H^{2}\;.
\ee

In the $\Lda$CDM paradigm,  current observations seem to suggest that the Friedmann equation is very well constrained by the relation
\eref{fried}
\be
H^{2}=\sfrac{1}{3}\left(\mu_d+\Lda\right)\;,
\label{hz}
\ee
where $\mu_d\ge0$ is the energy density of pressureless matter (dust) and $\Lambda$ is the cosmological constant. Planck's  most recent results \cite{planckpar} give a lower bound of the current value of the Hubble parameter at
\be
H_0=67.3\pm1.2\mbox{km}s^{-1}Mpc^{-1} \;,
\ee
as well as the current fractional energy densities of dust (cold dark matter plus baryonic matter) and dark energy (assumed to be due to the cosmological constant ) at
\be\label{eds}
\Omega_{0d}\equiv\frac{\mu_{0d}}{3H^{2}_{0}}=0.317\;,~~~~\Omega_{\Lambda}\equiv\frac{\Lambda}{3H^{2}_{0}}=0.683\;.
\ee
From Eqn (\ref{hz}) we see that the first and second time derivatives of the
scale factor $a(t)$ can be given as \cite{dunsby10}
\be\label{adots}
\dot{a}=\sqrt{\frac{\mu_{0d}}{3a}+\frac{\Lambda}{3}}\;;~~~~~\ddot{a}=\frac{1}{2}(\dot{a}^2)_{,a}=\frac{2\Lambda a^3-\mu_{0d}}{6a^2}\;.
\ee

Using these relations  in Eqn (\ref{R1}) one can show the following interrelation between the scale factor, the Ricci scalar and their  derivatives:
\ber\label{aR}
&&R(a)=\frac{4\Lambda a^3+\mu_0}{a^3}\;,~~~~~~~~~~~~~~~a(R)=\left(\frac{\mu_{0d}}{R-4\Lambda}\right)^{(1/3)}\;,\\
&&\label{HRdot} H(R)=\frac{1}{a(R)}\sqrt{\frac{\mu_0}{3a(R)}+\frac{\Lambda}{3}}\;,~~~~\dot{R}=R_{,a}(a(R))\sqrt{\frac{\mu_0}{3a(R)}+\frac{\Lambda}{3}}\;.
\eer

If we substitute all the above background quantities (expressed in terms of the Ricci scalar) into the Friedmann equation \eref{fried}, we obtain 
\be
6(R-3\Lambda)(R-4\Lambda)f''-(R-6\Lda)f'-f+2\mu_{m}=0\;,
\ee
the solution of which mimics the $\Lda$CDM expansion history exactly.

Let us now consider matter comprising dust-like matter plus a  non-interacting stiff fluid (or a massless scalar field) , i.e., 
\be
\mu_{m}=\mu_{d}+\mu_{s}\;,
\ee
with their present energy densities  $\mu_{0d}$ and $\mu_{0s}$ and  barotropic equations of state given by $w_{d}=0$ and $w_{s}=1$, respectively.  

The energy and momentum conservation equations for the two\hs fluid system 
in the energy frame  of matter (where the four-velocity vector field is aligned with the 
four-velocity of the total matter, i.e.,  $u^{a}=u^{a}_{m}$) is given as
\ber
&&\dot{\mu}_{d}+\mu_{d}\Theta =0\;,\\ 
&&\dot{\mu}_{s}+2\Theta\mu_{s}=0\;.\label{flux}
\eer

From the conservation equations, the total matter density is
\be
\mu_{m}(a)=\frac{\mu_{0d}}{a^3}+\frac{\mu_{0s}}{a^6}\Rightarrow\mu_{m}(R)=(R-4\Lambda)
+\frac{\mu_{0s}}{\mu_{0d}^2}(R-4\Lambda)^2\;.
\ee
Substituting this into the Friedmann equation (\ref{fried}), we get the following
particular solution \cite{dunsby10}:
\be
f(R)=R+\alpha R^2-2\Lambda\;,
\label{theory}
\ee
where 
\be\label{alpha}
\alpha=-\frac{2}{9}\frac{\mu_{0s}}{\mu^{2}_{0d}}=-\frac{2\Omega_{0s}}{27H^2_{0}\Omega^2_{0d}}\;.
\ee
where $\Omega_{0s}\equiv\frac{\mu_{0s}}{3H^{2}_{0}}$.

Thus the theory of gravity described by (\ref{theory}) has an exact 
$\Lambda CDM$ solution in the background level for a non\hs interacting two\hs fluid system of 
dust and a stiff matter like a massless scalar field. However, it is interesting to note that 
$\alpha<0$ if both the dust and scalar field densities are positive. Such theories 
have {\it ghosts} which exactly compensate for the massless scalar field and this is the 
reason that the solution does not depend on the scalar field dynamics. 
On the other hand, if we consider a massless {\it ghost} field with the dust, then 
$\alpha>0$ and the ghost field compensates for the extra degrees of freedom of the 
fourth\hs order gravity.

%%%%%%%%%%%%%%%%%%%%%%%%%%%%%%%%%%%%%%

%%%%%%%%%%%%%%%%%%%%%%%%%%%%%%%%%%%%%%

\section{Dynamics of scalar perturbations of the two-fluid system}\label{dynamics}

 In what follows, we investigate the stability of the background described  in the 
previous section with respect to generic linear 
inhomogeneous and anisotropic perturbations by linearizing the most general 
propagation and constraint equations for this $f(R)$ theory around the $\Lambda CDM$ solution. 
This is done using the 
$1+3$\hs covariant approach to perturbations \cite{EB89,BDE92,DBE92,dunsby91,carloni08,DBBE99,ananda2008}, where 
quantities that vanish 
in the background spacetime are considered to be first order and are 
automatically gauge\hs invariant locally by virtue of the 
Stewart\hs Walker lemma \cite{SW74}. 
In order to write down the set of linear perturbation evolution equations we first 
need to choose a physically motivated frame $u^a$. 
The natural choice is the one which is along the four-velocity of the total matter.

The usual covariant and gauge\hs invariant inhomogeneity variables of the total matter and expansion are given by \cite{DBE92, EB89}

\ber\label{gradmatt}
&&\md^{m}_{a}=\frac{a\D_{a}\mu_{m}}{\mu_{m}}\;,~~~~~~~\mz_{a}=a\D_{a}\Theta\;,\end{eqnarray}
whereas the information about our deviation from standard GR is carried by the following  dimensionless gradient quantities  \cite{carloni08, carlonireview}:
\be\label{gradcur}
{\car}_{a}=a\D_{a}R\;,~~~~~~ \Re_{a}=a\D_{a}\dot{R}\;.
\ee
For the dust and stiff-component fluids, we have
\be
\md^{d}_{a}=\frac{a\D_{a}\mu_{d}}{\mu_{d}}\;,~~~ \md^{s}_{a}=\frac{a\D_{a}\mu_{s}}{\mu_{s}}
\ee
where $\mu_{m}=\mu_{d}+\mu_{s}$ and $\mu_{m}\md^{m}_{a}=\mu_{d} \md^{d}_{a}+\mu_{s}\md^{s}_{a}\;.$
The  set of linearized evolution equations describing the perturbations in the  dust\hs stiff fluid mixture  is given by the  first-order coupled system of equations \cite{abebe12, carloni08}:
\ber\label{deldota}
&&\dot{\md}^{d}_{a}+\mz_{a}=0\;,\\
&&\dot{\md}^{s}_{a}-\Theta\md^{s}_{a}+2\mz_{a}=0\;,\\
&&\label{zdota} \dot{\mz}_{a}-\left(\dot{R}\frac{f''}{f'}-\frac{2}{3}\Theta\right)\mz_{a}+\frac{\mu_{m}}{f'}\md^{m}_{a}-\Theta\frac{f''}{f'}\Re_{a}+\frac{f''}{f'}\tl \nb^{2}{\car}_{a}\nn
&&~~~-\left( \frac{1}{2}-\frac{1}{2}\frac{ff''}{f'^{2}}+\frac{f''\mu_{m}}{f'^{2}}-\dot{R}\Theta(\frac{f''}{f'})^{2}+\dot{R}\Theta \frac{f'''}{f'}\right) {\car}_{a}=0\;,\\
&&\label{cardota}\dot{\car}_{a}-\Re_{a}=0\;,\\
&&\label{redota}\dot{\Re}_{a}+\left(\Theta +2\dot{R}\frac{f'''}{f''}\right)\Re_{a}+\dot{R}\mz_{a}-\frac{\mu_{m}}{3f''}\md^{m}_{a}-\tl\nb^{2}\car_{a}\nn
&&~~~+\left(\ddot{R}\frac{f'''}{f''}+\dot{R}^{2}\frac{f^{(4)}}{f''}+\Theta\dot{R}\frac{f'''}{f''}+\frac{f'}{3f''}-\frac{R}{3}\right)\car_{a}=0\;.
\eer
This system of 5-coupled partial differential equations involving vectorial gradients is difficult to solve. However, using the technique of uniquely decompsoing a vector into  divergence-free (solenoidal) and  curl-free (irrotational) parts and applying harmonic decomposition,  the above system can be rendered easily solvable.
\subsection{Evolution of the scalar perturbations}
If we take the spherically symmetric (trace) of gradient quantities 
(vectors) defined in \eref{gradmatt} and \eref{gradcur}, we obtain variables that characterize the evolution of the spherically 
symmetric part of the gradients. Since matter on cosmological scales is generally thought to follow 
spherical clustering, these new variables, given below, are what we are interested in 
knowing the evolutions of:
\ber
&&\del_{m}=a\tl\nb^{a}\md^{m}_{a}\;, ~~~~~  \del_{d}=a\tl\nb^{a}\md^{d}_{a}\;, ~~~~~ \del_{s}=a\tl\nb^{a}\md^{s}_{a}\;, \nonumber \\
&&\mz=a\tl\nb^{a}\mz_{a}\;,~~~~~~~~\car=a\tl\nb^{a}{\car}_{a}\;,~~~~~~  \Re=a\tl\nb^{a}\Re_{a}\;.
\eea

Given below are the evolution equations of the {\it harmonically decomposed} (see \cite{BDE92} for a detailed treatment of harmonics) gradient variables:
\ber
&&\dot\del^{\k}_{d}+\mz^{\k}=0\;,\\
&&\dot\del^{\k}_{s}-\Theta\del_{s}+2\mz^{\k}=0\;,\\
&&\dot \mz^{\k}-(\dot{R}\frac{f''}{f'}-\frac{2}{3}\Theta)\mz^{\k}+\frac{\mu_{m}}{f'} \del^{\k}_{m}-\Theta\frac{f''}{f'}\Re^{\k}\nonumber\\
&&~~~-\left[ \frac{1}{2}+\frac{f''}{f'}\frac{\k^{2}}{a^{2}}-\frac{1}{2}\frac{ff''}{f'^{2}}+\frac{f''\mu_{m}}{f'^{2}}-\dot{R}\Theta(\frac{f''}{f'})^{2}+\dot{R}\Theta \frac{f'''}{f'}\right] {\car}^{\k}=0\;,\\
&&\dot{\car}^{\k}-\Re^{\k}=0\;,\\
&&\dot\Re^{\k}+\left(\Theta +2\dot{R}\frac{f'''}{f''}\right)\Re^{\k}+\dot{R}\mz^{\k}-\frac{\mu_{m}}{3f''}\del^{\k}_{m}\nonumber\\
&&~~~+\left[\frac{\k^{2}}{a^{2}}+\ddot{R}\frac{f'''}{f''}+\dot{R}^{2}\frac{f^{(4)}}{f''}+\Theta\dot{R}\frac{f'''}{f''}+\frac{f'}{3f''}-\frac{R}{3}\right]\car^{\k}=0\;.
\eer
where any separable scalar gradient is defined in terms of its harmonic components by
\be
X=\sum_{k}{X^{\k}(t)}Q_{k}(x)
\ee
and using the Laplace-Beltrami operator
\be 
\tlnb Q=-\frac{\k^{2}}{a^{2}}Q
\ee
with the wave number $\k=\frac{2\pi a}{\lambda}$ and $\dot{Q}_{x}=0$.

In redshift space, the evolutions of the perturbations can be  given by

\begin{eqnarray}\label{dd}
&&(1+z)^{2}\dot{a}\frac{d\del^{\k}_{d}}{dz}=\mz^{\k},\\
&&(1+z)^{2}\dot{a}\frac{d\del^{\k}_{s}}{dz}=-\Theta\del^{\k}_{s}+2\mz^{\k}\;,\\
&&\label{ezed}(1+z)^{2}\dot{a}\frac{d\mz^{\k}}{dz}=-(\dot{R}\frac{f''}{f'}-\frac{2}{3}\Theta)\mz^{\k}+\frac{\mu_{m}}{f'} \del^{\k}_{m}-\Theta\frac{f''}{f'}\Re^{\k}\nonumber\\
&&~~~-\left[ \frac{1}{2}+\frac{f''}{f'}\frac{\k^{2}}{a^{2}}-\frac{1}{2}\frac{ff''}{f'^{2}}+\frac{f''\mu_{m}}{f'^{2}}-\dot{R}\Theta(\frac{f''}{f'})^{2}+\dot{R}\Theta \frac{f'''}{f'}\right] {\car}^{\k},\\
&&(1+z)^{2}\dot{a}\frac{d{\car}^{\k}}{dz}=-\Re^{\k},\\
&&\label{resz}(1+z)^{2}\dot{a}\frac{d\Re^{\k}}{dz}=\left(\Theta +2\dot{R}\frac{f'''}{f''}\right)\Re^{\k}+\dot{R}\mz^{\k}-\frac{\mu_{m}}{3f''}\del^{\k}_{m}\nonumber\\
&&~~~+\left[\frac{\k^{2}}{a^{2}}+\ddot{R}\frac{f'''}{f''}+\dot{R}^{2}\frac{f^{(4)}}{f''}+\Theta\dot{R}\frac{f'''}{f''}+\frac{f'}{3f''}-\frac{R}{3}\right]\car^{\k}\;,
\end{eqnarray}
where the usual definition of redshift $1+z=\sfrac{a_{0}}{a}$ is used\footnote{For practical purposes, we have normalized $a_{0}=1$ today.}.

To solve the set of scalar perturbation equations numerically, we redefine the following normalized quantities: 
\ber
&&\mz(z)=H_{0}\mz_{n}(z)\;,~~~
{\car(z)}=H^{2}_{0}{\car_{n}(z)}\;,~~~~~~\Re(z)=H^{3}_{0}\Re_{n}(z)\nn
&&H(z)=H_{0}h(z)\;,~~~~~~~\k=H_{0}k\;.\eer
Using the above quantities 
 we are able to rewrite Equations (\ref{dd}-\ref{resz}) as a system of five ODEs:
 
\begin{eqnarray}\label{eq1}
&&(1+z)\sqrt{(1+z)^{3}\Omega_{0d}+\Omega_{\Lambda}}\frac{d\del^{\k}_{d}}{dz}=\mz^{\k}_{n}(z),\\
&&(1+z)\sqrt{(1+z)^{3}\Omega_{0d}+\Omega_{\Lambda}}\frac{d\del^{\k}_{s}}{dz}=-3\sqrt{(1+z)^{3}\Omega_{0d}+\Omega_{\Lambda}}\del^{\k}_{s}+2\mz^{\k}_{n}(z)\;,\\
%%%%%%%%%%%%%%%%%%%%%%%%%%%
&&(1+z)\sqrt{(1+z)^{3}\Omega_{0d}+\Omega_{\Lambda}}\frac{d\mz^{\k}_{n}(z)}{dz}=\left[2\sqrt{(1+z)^{3}\Omega_{0d}+\Omega_{\Lambda}}-\frac{36\Omega_{0d}\Om_{0s}(1+z)^{3}}{27\Om^{2}_{0d}-4\Om_{0s}}\right]\mz^{\k}_{n}(z)\nn
&&~~~~~~+\frac{3(1+z)^{3}\left[\Omega_{0d}\del^{\k}_{d}+(1+z)^{3}\Omega_{0s}\del^{\k}_{s}\right]}{\left[1-\frac{4\Om_{0s}}{9\Om^{2}_{0d}} \left((1+z)^{3}\Omega_{0d}+4\Omega_{\Lambda}\right)\right]} +\frac{4\Om_{0s}}{9\Om^{2}_{0d}-4\Om_{0s} \left((1+z)^{3}\Omega_{0d}+4\Omega_{\Lambda}\right)}\Re^{\k}_{n}\nn
&&~~~~~-\left[ \frac{1}{2}-\frac{4\Om_{0s}(1+z)^{2}k^{2}}{3\left[9\Om^{2}_{0d}-4\Om_{0s} \left((1+z)^{3}\Omega_{0d}+4\Omega_{\Lambda}\right)\right]}\right.\nn
&&~~~~~\left.+\frac{2\Om_{s}\left[3(1+z)^{3}\Omega_{0d}+6\Omega_{\Lambda}-\frac{2\Om_{s}\left((1+z)^{3}\Omega_{0d}+4\Omega_{\Lambda}\right)^{2}}{3\Om^{2}_{0d}}\right]}{27\Om^{2}_{0d}\left[1-\frac{4\Om_{0s}}{9\Om^{2}_{0d}} \left((1+z)^{3}\Omega_{0d}+4\Omega_{\Lambda}\right)\right]^{2}}-\frac{4\Om_{s}\left[3(1+z)^{3}\left(\Omega_{0d}+(1+z)^{3}\Om_{0s}\right)\right]}{27\Om^{2}_{0d}\left[1-\frac{4\Om_{0s}}{9\Om^{2}_{0d}} \left((1+z)^{3}\Omega_{0d}+4\Omega_{\Lambda}\right)\right]^{2}}\right.\nn
&&~~~~~\left.+\frac{16\Om_{0s}^{2}\left[(1+z)^{3}\left((1+z)^{3}\Om_{0d}+\Om_{\Lda}\right)\right]}{27\Om_{0d}^{3}[1-\frac{4\Om_{0s}}{9\Om^{2}_{0d}} \left((1+z)^{3}\Omega_{0d}+4\Omega_{\Lambda}\right)]^{2}}\right]{\car}^{\k}_{n}=0\;,\\
%%%%%%%%%%%%%%%%%%%%%%%%%%%%%%
&&(1+z)\sqrt{(1+z)^{3}\Omega_{0d}+\Omega_{\Lambda}}\frac{d{\car}^{\k}_{n}}{dz}=-\Re^{\k}_{n}\;,\\
&&\label{eq5}(1+z)\sqrt{(1+z)^{3}\Omega_{0d}+\Omega_{\Lambda}}\frac{d\Re_{n}^{\k}}{dz}=3\sqrt{(1+z)^{3}\Omega_{0d}+\Omega_{\Lambda}}\Re_{n}^{\k}-9\Omega_{0d}(1+z)^{3}h(z)\mz^{\k}_{n}\nn
&&~~~~~+\frac{81\Om^{2}_{0d}(1+z)^{3}\left[\Om_{od}\del^{\k}_{d}+\Om_{os}(1+z)^{3}\del^{\k}_{s}\right]}{4\Omega_{0s}}\nn
&&~~~~~+\left[(1+z)^{2}k^{2}-\frac{{27\Om^{2}_{0d}}-4\Om_{0s}\left[(1+z)^{3}\Om_{0d}+4\Om_{\Lambda}\right]}{4\Om_{0s}}-(1+z)^{3}\Om_{od}-4\Om_{\Lambda}\right]\car_{n}^{\k}.\nn
\end{eqnarray}

%%%%%%%%%%%%%%%%%%%%%%%%%%%%%%%%%%%%%%%%%%%%%%%%%%%%%%%%%%%%%%%%%%%%%%%%%%%%%%%%%%%%%
We solve the above evolution equations setting the initial fluctuations at $z=1$ to be $\sim 10^{-5}$ and using the fractional energy densitities of dust and the cosmological constant given by Eqn \eref{eds}.

The $k$\hs dependence of the amplitudes of  the perturbations at a given redshift is depicted by the plots of the  power spectrum defined as \cite{ananda2009}:
\be
\langle \del_{d}({\bf{k}}_{1})\del_{d}({\bf{k}}_{2})\rangle=P(k_{1})\delta(\bf{k}_{1}+\bf{k}_{2}),
\ee
where ${\bf{k}}_{1}$ and ${\bf{k}}_{2}$ are [normalized] wavevectors of Fourier components of the solutions of the above system.  Since  isotropy of the perturbations is assumed, we can simply write $P(k_{1})$ instead of $P(\bf{k}_{1})\;.$

The power spectrum of the fluctuations in GR (plus  $\Lda$CDM) is scale-invariant. Thus any deviation from such invariance in our $f(R)$ models calls for a closer analysis of structure formation scenarios in such models. It is also a powerful way of putting tight constraints on the viability of the $f(R)$ gravitational models in question. 

\begin{figure}[!htb]
\minipage{0.32\textwidth}
  \includegraphics[width=\linewidth]{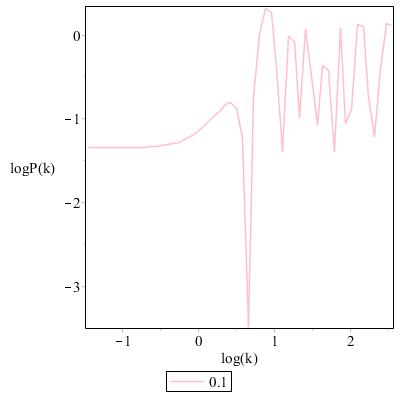}
  \caption{Power spectrum for  $\Omega_{0s}=10^{-1}$.}\label{spec1}
\endminipage\hfill
\minipage{0.32\textwidth}
  \includegraphics[width=\linewidth]{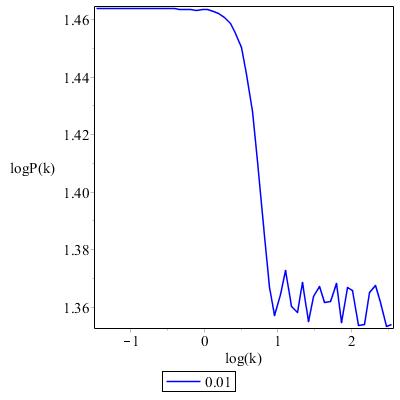}
  \caption{Power spectrum for  $\Omega_{0s}=10^{-2}$.}\label{spec2}
\endminipage\hfill
\minipage{0.32\textwidth}%
  \includegraphics[width=\linewidth]{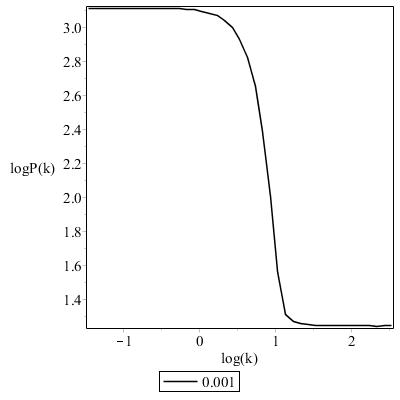}
 \caption{Power spectrum for  $\Omega_{0s}=10^{-3}$.}\label{spec3}
\endminipage
\end{figure}

\begin{figure}[!htb]
\minipage{0.32\textwidth}
  \includegraphics[width=\linewidth]{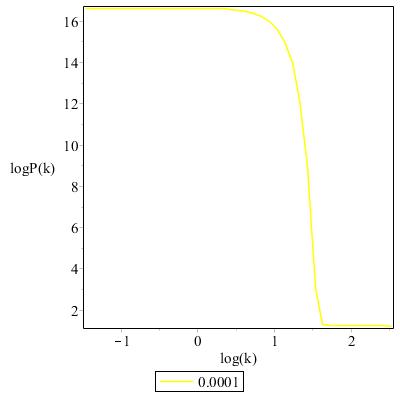}
  \caption{Power spectrum for  $\Omega_{0s}=10^{-4}$.}\label{spec4}
\endminipage\hfill
\minipage{0.32\textwidth}
  \includegraphics[width=\linewidth]{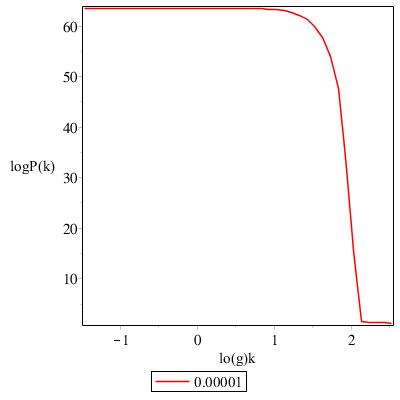}
  \caption{Power spectrum for  $\Omega_{0s}=10^{-5}$.}\label{spec5}
\endminipage\hfill
\minipage{0.32\textwidth}%
  \includegraphics[width=\linewidth]{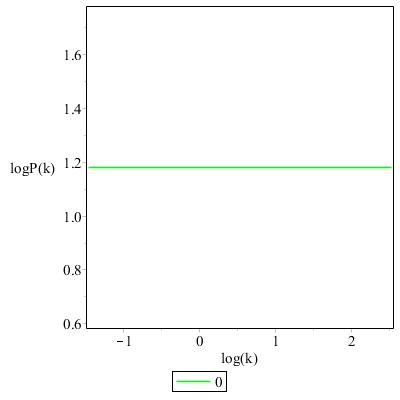}
 \caption{Power spectrum for  $\Omega_{0s}=0$. Note that $\Om_{0s}=0$ corresponds to GR+$\Lda$CDM; hence the scale-invariance.}\label{specg}
\endminipage
\end{figure}

\begin{figure}[here!]
\begin{center}
\includegraphics[scale=.55]{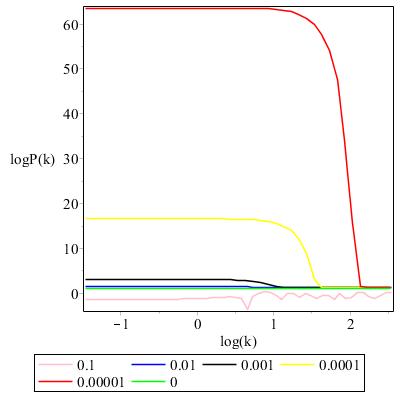}\\
\caption{Power spectrum for different values of  $\Omega_{s}=0\;,10^{-5}\;, 10^{-4}\;, 10^{-3}\;, 10^{-2}\;, 10^{-1}$.}\label{specall}
\end{center}
\end{figure}

%%%%%%%%%%%%%%%%%%%%%%%%%%%%%%%%%%%%%%%%%%%%%%%%%%%%%%%%%%%%%%%%%%%%%%%%%%%%%%%%%%%%%
The following plots (Figs. \eref{spec1}-\eref{specall}) show how the growth of the dust perturbations depend on scale for different values of $\Om_{s}$ when $\alpha<0$. Here we are interested in the relative magnitudes of the power spectra and  have therefore plotted the  normalized power spectral values. The vertical axis shows the logarithm of the normalized power spectrum $P(k)=\left[\frac{\Delta_{d}(k)}{\Delta_{d}(k=10, \Om_{s}=0.1)}\right]^{2}$ today $(z=0)$ whereas the horizontal axis is the logarithm of $k$.  Note the general drop in power for the $R+\alpha R^{2}-2\Lda$ model as opposed to the scale indifference of the perturbations in GR.
\np
We see from these figures  that, even though the models of gravity studied govern identical background dynamics, this degeneracy has been broken at first-order perturbations level.

As can be seen in the plots, the smallest values of k for which the power spectrum  starts to deviate from flatness depend on how far our model is from $\Lambda$CDM. For example, for positive $\Omega_{0s}$ values,  we can observe from the plots that if we are very close to $\Lambda$CDM, i.e., if $\Omega_{0s}$ values are smaller), then the deviation from flatness occurs at larger k values ($1<log(k)<2 $ for $\Omega_{0s}=10^{-5}$, $0<log(k)<1$ for $\Omega_{0s}=10^{-4}$,  $-1<log(k)<0 $ for $\Omega_{0s}=0.1$, etc.)

\begin{figure}[!htb]
\minipage{0.32\textwidth}
  \includegraphics[width=\linewidth]{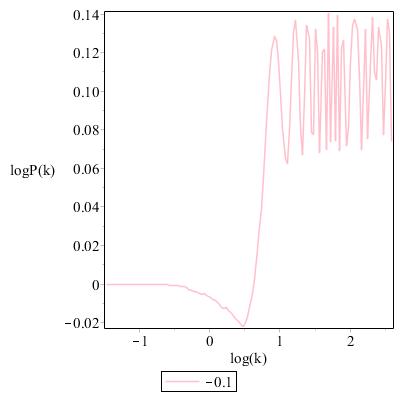}
  \caption{Power spectrum for  $\Omega_{0s}=-10^{-1}$.}\label{specn1}
\endminipage\hfill
\minipage{0.32\textwidth}
  \includegraphics[width=\linewidth]{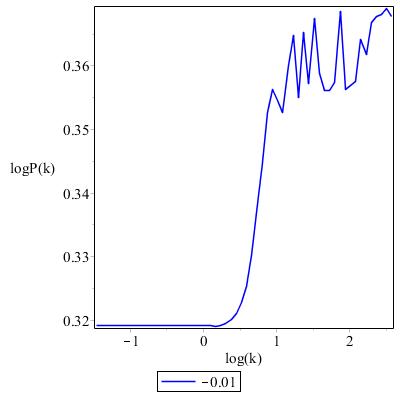}
  \caption{Power spectrum for  $\Omega_{0s}=-10^{-2}$.}\label{specn2}
\endminipage\hfill
\minipage{0.32\textwidth}%
  \includegraphics[width=\linewidth]{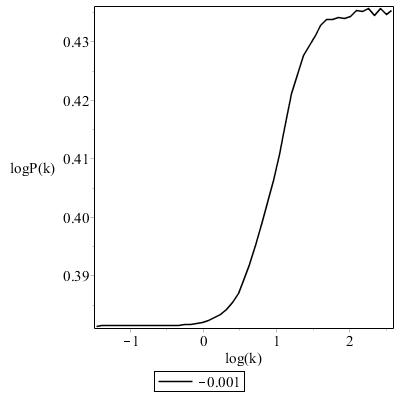}
 \caption{Power spectrum for  $\Omega_{0s}=-10^{-3}$.}\label{specn3}
\endminipage
\end{figure}

\begin{figure}[!htb]
\minipage{0.32\textwidth}
  \includegraphics[width=\linewidth]{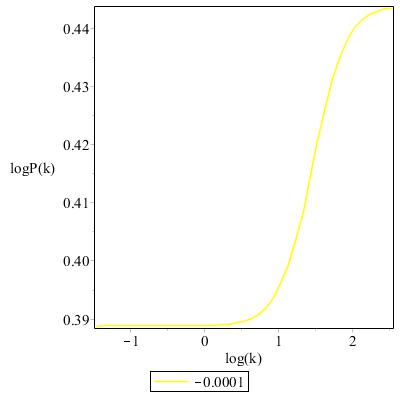}
  \caption{Power spectrum for  $\Omega_{0s}=-10^{-4}$.}\label{specn4}
\endminipage\hfill
\minipage{0.32\textwidth}
  \includegraphics[width=\linewidth]{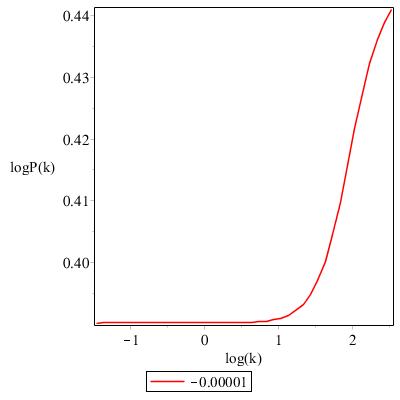}
  \caption{Power spectrum for  $\Omega_{0s}=-10^{-5}$.}\label{specn5}
\endminipage\hfill
\minipage{0.32\textwidth}%
  \includegraphics[width=\linewidth]{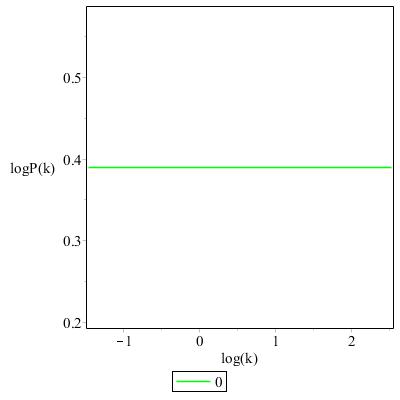}
   \caption{Power spectrum for  $\Omega_{0s}=0$.}\label{specgn}
\label{spectralln}
\endminipage
\end{figure}

\begin{figure}[here!]
\begin{center}
\includegraphics[scale=.55]{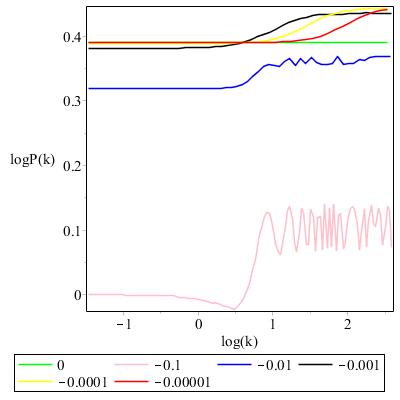}\\
\caption{Power spectrum for different values of  $\Omega_{0s}=-10^{-1}\;, -10^{-2}\;, -10^{-3}\;,$ $-10^{-4}\;, -10^{-5}\;,0\;$. }\label{specalln}
\end{center}
\end{figure}

In the case of $\alpha>0$, we observe a flat spectrum (Figs. \eref{specn1}-\eref{specalln}) on the longer scale regimes and the spectra start to rise on smaller and smaller scales (larger and larger $\kappa$ values). The normalized power spectra at $z=0$ are defined for this particular case as  $P(k)=\left[\frac{\Delta_{d}(k)}{\Delta_{d}(k=10, \Om_{s}=-0.1)}\right]^{2}$.

We also notice a collapse of the background  degeneracy for these models. The important lesson here is that it is not sufficient to just study the gravitational model that describes the cosmological background expansion history and  conclude that that is the right description of gravitational physics.

\section{The quasi\hs static approximation}\label{quasi}
It has been shown \cite{abebe12, bertschinger08, abebe13} that on small  scales, the contributions from the temporal fluctuations of 
$\car^{\k}$ can be neglected, 
i.e., $\dot{\car}^{\k}\simeq 0\Rightarrow \Re^{\k}\simeq 0 $, 
because they are quickly damped away. If we apply this approximation to Eqns \eref{eq1}-\eref{eq5} the resulting sub-horizon ($\kappa^{2}\gg H^{2}$) perturbation equations reduce to
\ber
&&(1+z)\sqrt{(1+z)^{3}\Omega_{0d}+\Omega_{\Lambda}}\frac{d\del^{\k}_{d}}{dz}=\mz^{\k}_{n}(z),\\
&&(1+z)\sqrt{(1+z)^{3}\Omega_{0d}+\Omega_{\Lambda}}\frac{d\del^{\k}_{s}}{dz}=-3\sqrt{(1+z)^{3}\Omega_{0d}+\Omega_{\Lambda}}\del^{\k}_{s}+2\mz^{\k}_{n}(z)\;,
\eer
\ber
%%%%%%%%%%%%%%%%%%%%%%%%%%%
&&(1+z)\sqrt{(1+z)^{3}\Omega_{0d}+\Omega_{\Lambda}}\frac{d\mz^{\k}_{n}(z)}{dz}=\left[2\sqrt{(1+z)^{3}\Omega_{0d}+\Omega_{\Lambda}}-\frac{36\Omega_{0d}\Om_{0s}(1+z)^{3}}{27\Om^{2}_{0d}-4\Om_{0s}}\right]\mz^{\k}_{n}(z)\nn
&&~~~~~~+\frac{3(1+z)^{3}\left[\Omega_{0d}\del^{\k}_{d}+(1+z)^{3}\Omega_{0s}\del^{\k}_{s}\right]}{\left[1-\frac{4\Om_{0s}}{9\Om^{2}_{0d}} \left((1+z)^{3}\Omega_{0d}+4\Omega_{\Lambda}\right)\right]}-\left[ \frac{1}{2}-\frac{4\Om_{0s}(1+z)^{2}k^{2}}{3\left[9\Om^{2}_{0d}-4\Om_{0s} \left((1+z)^{3}\Omega_{0d}+4\Omega_{\Lambda}\right)\right]}\right.\nn
&&~~~~~\left.+\frac{2\Om_{s}\left[3(1+z)^{3}\Omega_{0d}+6\Omega_{\Lambda}-\frac{2\Om_{s}\left((1+z)^{3}\Omega_{0d}+4\Omega_{\Lambda}\right)^{2}}{3\Om^{2}_{0d}}\right]}{27\Om^{2}_{0d}\left[1-\frac{4\Om_{0s}}{9\Om^{2}_{0d}} \left((1+z)^{3}\Omega_{0d}+4\Omega_{\Lambda}\right)\right]^{2}}-\frac{4\Om_{s}\left[3(1+z)^{3}\left(\Omega_{0d}+(1+z)^{3}\Om_{0s}\right)\right]}{27\Om^{2}_{0d}\left[1-\frac{4\Om_{0s}}{9\Om^{2}_{0d}} \left((1+z)^{3}\Omega_{0d}+4\Omega_{\Lambda}\right)\right]^{2}}\right.\nn
&&~~~~~\left.+\frac{16\Om_{0s}^{2}\left[(1+z)^{3}\left((1+z)^{3}\Om_{0d}+\Om_{\Lda}\right)\right]}{27\Om_{0d}^{3}[1-\frac{4\Om_{0s}}{9\Om^{2}_{0d}} \left((1+z)^{3}\Omega_{0d}+4\Omega_{\Lambda}\right)]^{2}}\right]\times \left[\frac{9\Om_{0d}(1+z)^{3}\sqrt{(1+z)^{3}\Omega_{0d}+\Omega_{\Lambda}}}{4\Om^{2}_{0s}(1+z)^{2}k^{2}-9\Om^{2}_{0d}}\mz^{\k}_{n}(z)\right.\nn
&&\left.~~~~~-\frac{81\Om_{0d}^{2}(1+z)^{3}\left[\Om_{0d}\del^{\k}_{d}+\Om^{2}_{0s}(1+z)^{3}\del^{\k}_{s}\right]}{4\Om^{2}_{0s}(1+z)^{2}k^{2}-9\Om^{2}_{0d}} \right]\;.
\eer
%%%%%%%%%%%%%%%%%%%%%%%%%%%%%%
 One can see from the following plots that the quasi\hs static solutions (blue dashed lines) of the dust perturbations are indeed a good approximation to the  solution of the full equations (red solid lines) for the $R+\alpha R^{2}-2\Lda$ model. In these plots, the vertical axis $\Delta_{d}(z)\equiv \Delta$ corresponds to the amplitudes of the dust perturbations. 

\begin{figure}[!htb]
\minipage{0.32\textwidth}
  \includegraphics[width=\linewidth]{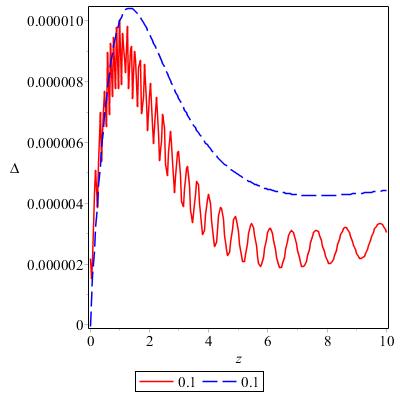}
  \caption{Dust density perturbations  for $\Omega_{s}=10^{-1}$ at $\kappa=10^{2}$ in the quasi\hs static approximation.}\label{delfq1}
\endminipage\hfill
\minipage{0.32\textwidth}
  \includegraphics[width=\linewidth]{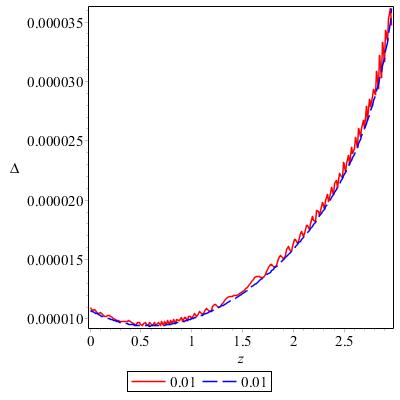}
  \caption{Dust density perturbations  for $\Omega_{s}=10^{-2}$ at $\kappa=10^{3}$ in the quasi\hs static approximation.}\label{delfq2}
\endminipage\hfill
\minipage{0.32\textwidth}%
  \includegraphics[width=\linewidth]{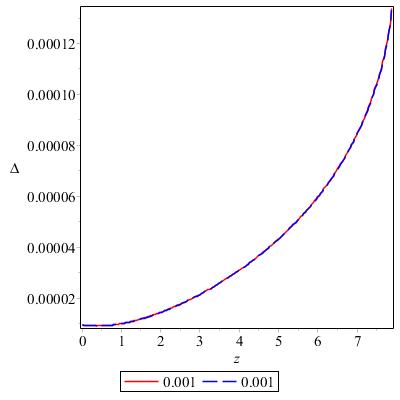}
  \caption{Dust density perturbations  for $\Omega_{s}=10^{-3}$ at $\kappa=10^{4}$ in the quasi\hs static approximation.}\label{delfq3}
\endminipage
\end{figure}

\begin{figure}[!htb]
\minipage{0.32\textwidth}
  \includegraphics[width=\linewidth]{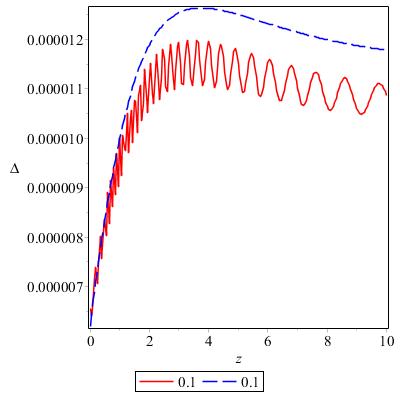}
  \caption{Dust density perturbations  for $\Omega_{s}=-10^{-1}$ at $\kappa=10^{2}$ in the quasi\hs static approximation.}\label{delfqn1}
\endminipage\hfill
\minipage{0.32\textwidth}
  \includegraphics[width=\linewidth]{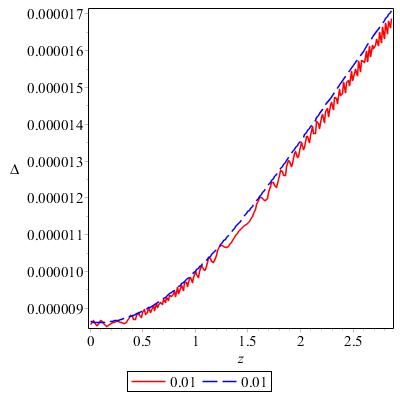}
  \caption{Dust density perturbations  for $\Omega_{s}=-10^{-2}$ at $\kappa=10^{3}$ in the quasi\hs static approximation.}\label{delfqn2}
\endminipage\hfill
\minipage{0.32\textwidth}%
  \includegraphics[width=\linewidth]{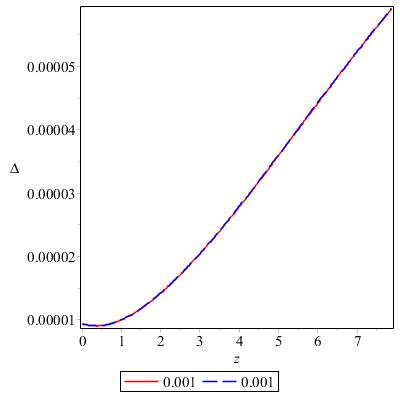}
  \caption{Dust density perturbations  for $\Omega_{s}=-10^{-3}$ at $\kappa=10^{4}$ in the quasi\hs static approximation.}\label{delfqn3}
\endminipage
\end{figure}

We observe from these plots (Figs.\eref{delfq1}-\eref{delfqn3})  that for large enough $\kappa$ values (smaller scales), the quasi\hs static and full solutions are indistinguishable both for $\alpha<0$ ($\Om_{0s}>0$) and $\alpha>0$ ($\Om_{0s}<0$), thus confirming the validity of the quasi\hs static approximation for our present model.
\np
\section{Discussion and Conclusion}\label{disc}
Using the scheme of first\hs order covariant perturbations, we have shown that it is possible to break the degeneracy that exists between $\Lambda$CDM and the reconstructed $f(R)$  models which, at the background level, describe exactly the same cosmic expansion history. Using different values for the characteristic parameter $\alpha$ (or correspondingly different $\Om_{0s}$ by virtue of Eqn \eref{alpha}), we  have shown that first\hs order perturbations show that structures in these two (i.e., $\Lda$CDM and $R+\alpha R^{^2}-2\Lambda$) models evolve at different rates, the former independently of the wavenumber $\kappa$, the latter in accordance with the $\kappa$\hs dependence of Eqns \eref{ezed} and \eref{resz}. In the  long\hs wavelength regime, we observe flat power spectrum in both models. Moreover, the peculiar drop in power in the short\hs wavelength regime in some $f(R)$ theories \cite{ananda2009, abebe13} is a feature we have observed in the physically interesting ($\alpha<0\;, \Om_{s}>0$) models we  considered. In the case of ghost solutions ($\alpha>0\;, \Om_{s}<0$), while the background degeneracy  still remains broken, the power spectra rise, rather than drop, with decreasing scale (increasing $\kappa$\hs value). We have made an application of the quasi\hs static technique of approximation for the solutions of the perturbation equations. The technique appears to be a very good approximation to the full  short\hs wavelength solutions for the models we studied.

A natural extension of this work will be to consider nonlinear contributions to the perturbations and see if the degeneracy is enhanced.

The breaking of background degeneracy at the level of linearised perturbations as well as the scales on which the degeneracy is sufficiently  broken before nonlinear physics becomes significant can be studied using  current and forthcoming cosmological probes. These  probes might include  those with stakes in the study of   large-scale structure formation and dark energy, \cite{kunz09,aviles11, metcalf98}, gravitational lensing and baryonic acoustic oscillations as well as modified gravity  \cite{fedeli14, springel06, boylan09, eisen05, schmidt11} (such as SDSS, WMAP, PLANCK, LSST, and EUCLID).

\section*{References}

\bibliography{@bibliography}
%\renewcommand{\bibname}{References}
%\nocite{*}%\bibliographystyle{revcompchem}
%\bibliographystyle{naturemag}
%\bibliographystyle{amsalpha} %% acm, naturemag, revcompchem
%\bibliographystyle{alpha}
%\bibliographystyle{unsrt}
%\bibliographystyle{apalike}
%\bibliographystyle{amsplain}
%\bibliographystyle{plain}
%\bibliographystyle{h-physrev3.bst}
%\bibliographystyle{amsplain}
%\bibliographystyle{abbrv}
%\bibliographystyle{apsrev}
\bibliographystyle{iopart-num}

\end{document}